\documentclass[showpacs,twocolumn,prb]{revtex4-1}

\usepackage{fancyhdr}
\usepackage{graphicx}
\usepackage{amsmath}
\makeatletter
\newcommand{\biggg}{\bBigg@{4}}
\makeatother
\usepackage{geometry}
\makeatletter
\def\ps@myPS{%
    \def\@oddfoot{\null\hfill\thepage}
    \def\@evenfoot{\thepage}%
    \def\@evenhead{\null\hfil\slshape\leftmark}%
    \def\@oddhead{{\slshape\rightmark}}}%
\makeatother

\begin{document}
\title{The double well potential in quantum mechanics: a simple, numerically exact formulation}
\author{V. Jelic and F. Marsiglio}
\affiliation{Department of Physics, University of Alberta, Edmonton, Alberta, Canada, T6G~2E1}

\begin{abstract}

The double well potential is arguably one of the most important potentials in quantum mechanics, because the solution contains the notion of a state as a linear superposition of `classical' states, a concept which has become very important in quantum information theory. It is therefore desirable to have solutions to simple double well potentials that are accessible to the undergraduate student. We describe a method for obtaining the numerically exact eigenenergies and eigenstates for such a model, along with the energies obtained through the Wentzel-Kramers-Brillouin (WKB) approximation. The exact solution is accessible with elementary mathematics, though numerical solutions are required. We also find that the WKB approximation is remarkably accurate, not just for the ground state, but for the excited states as well.
\end{abstract}

\date{\today} 
\maketitle

\section{introduction}

One of the most lasting impressions of quantum mechanics in the public consciousness is that of the so-called Schr\"odinger cat paradox.\cite{griffiths} This notion, that physical systems generally exist as a superposition of various states, lies at the heart of quantum mechanics, and is emphasized, for example, when one studies the time evolution of a system initially prepared in such a superposition. Well publicized paradigms, such as the Einstein-Podolsky-Rosen (EPR) paradox,\cite{einstein35}, and the more modern tests of Bell's inequalities,\cite{mermin85,current05} are all manifestations of this superposition of states.

A typical physics undergraduate student is likely to have been drawn to physics in the first place by popular accounts of these and other phenomena. Part of maintaining this interest throughout their coursework is achieving the right balance between, on one hand, learning the necessary mathematics and `standard' solutions, as presented in most textbooks, and, on the other hand, actively engaging in problems with stimulating and visual solutions that illustrate some of these exciting ideas. For example, Feynman used the ammonia
molecule to illustrate the principle of superposition of states in quantum 
mechanics.\cite{feynman65} In systems like ammonia there are (typically) two degenerate states; the system can reside in a superposition of both, and, most importantly, can tunnel from one to the other. Tunnelling is omnipresent in quantum mechanics, and is the key ingredient in modern applications such as solid state devices (e.g. the diode), solar cells, and microscopes.\cite{wiesendanger94}

The example of the ammonia molecule is a rich problem with many details one could focus on, but the purpose of this paper is to introduce the student to a more microscopic example. The simplest model for studying a superposition of nearly degenerate states is the one dimensional double well potential. This problem consists of a potential with two minima separated by a barrier. Model potentials of a double well are indeed treated in most textbooks, and even in many pedagogical-style articles.\cite{holstein87,chebotarev98,garg00,gea-banacloche02,dutt10,foot11}

The treatment of a double well potential in textbooks will vary from qualitative discussion (see, for example, Problem 2.47 in Griffiths,\cite{griffiths}) to an analytical solution (for example, Section 8.5 in Merzbacher,\cite{merzbacher} where, however, advanced mathematical steps are required). More often, the Wentzel-Kramers-Brillouin (WKB) approximation is used (see, for example, Problem 8.15 in Griffiths,\cite{griffiths} and Problem 7.2, and 8.10 in Merzbacher.\cite{merzbacher}). In this paper we will present a method which allows students 
to numerically solve these tunneling problems for themselves without complicated mathematics. This method necessarily requires the students to access numerical procedures for diagonalizing a matrix, and has been explained in some detail in Ref. [\onlinecite{marsiglio09}]. The philosophy behind this teaching practice is that students can utilize familiar packages like Mathematica, Maple, and Matlab to both solve these problems and gain a visual and more memorable understanding of the solutions. This is currently not the case in most undergraduate classes, but asking students to carry out similar problems as outlined below will aid toward this goal. 

We begin by defining the problem in the next section, and outlining the numerically exact solution, following the treatment of the harmonic oscillator problem in Ref. [\onlinecite{marsiglio09}]. The primary focus is the ground state splitting, but many other levels, including the wave functions, can be obtained. Our method of solution requires embedding the double well potential in an infinite well, whose presence has no effect on the ground state and low lying excited state solutions. We note that a variety of double well potentials can be studied this way; furthermore, there is no restriction on having an analytic form for the potential. This brings students to the point that very realistic potentials can be studied without further complication, bringing the classroom experience significantly closer to research problems. Indeed, one of the underlying motivations for developing a set of problems like these, suitable to undergraduate training, is that these same
methods are generally adopted in research practice.

To make contact with more approximate analytic solutions, we outline the `standard' WKB solution, defined by using connection formulas at linear turning points; here again, most authors have focused on the ground state splitting, but since we have numerically exact results we will examine the WKB solution for all the states. This necessarily requires three separate calculations, one for the low-lying states inside the potential well, one for the higher states, where the energy of the state exceeds the central barrier, and one for still higher energy states, where the vertical walls of the infinite well come into play. These latter states arise only because we have adopted the methodology of Ref. [\onlinecite{marsiglio09}], but the comparison is useful anyways for a proper assessment of the WKB approximation for all energy levels.\cite{remark2} This section is sufficiently detailed in part to illustrate that the WKB solution, when done fully, is actually very accurate. In the end, the WKB approximation, which is only valid for certain potentials, requires considerably more work than the numerically exact solution. 

The take-home message of this paper is that utilizing the matrix mechanics of Ref. [\onlinecite{marsiglio09}] places the entire problem of the double well potential in full control of undergraduate students, and prepares them for even more complicated problems.

\section{The double well potential: numerically exact solution}

Several analytic forms for the double well potential are available for consideration, and are illustrated in Fig.~1. For a square double well the potential is defined by
\begin{equation}
V_{\text sq}(x) = \begin{cases} 0 &
\text{if $| x -({a \over 2} - b)| <  {c \over 2}$ or $| x -({a \over 2} + b)| <  {c \over 2}$} \\ 
V_0 & \text{otherwise,}
\end{cases}
\label{square_doublewell_potential} 
\end{equation}
where the potential is centered around $x = a/2$, and the parameters $b$ and $c$ control the positioning and width of the individual wells, respectively. It is of course possible to formulate this problem analytically, but it is tedious, and more simply handled through the formalism described below.

\begin{figure}[h!]
\begin{center}
\includegraphics[height=3.5in,width=3.0in]{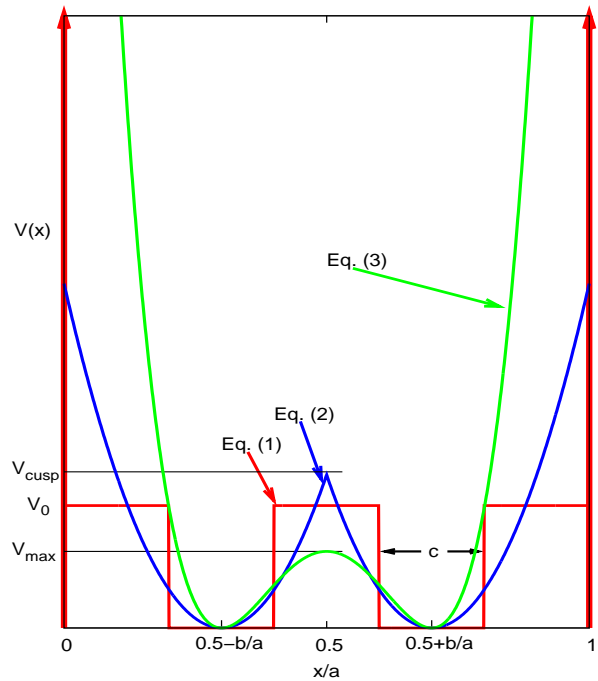} 
\caption{(color online) The various possible double well potentials used in this problem are shown as a function of $x$. 
All are centred around $x=a/2$, and all have minima centered around $x_m = a/2 \pm b$. The square double well potential is shown in red, with the analytic form given by Eq. \eqref{square_doublewell_potential}. The quadratic form is shown in blue, and is given by Eq. (\ref{potential}); it has a maximum (cusp) at the value $V_{\rm cusp}$. The quartic form is shown in green, and is described by Eq. \eqref{potential_2}; it has a smooth maximum at the value $V_{\rm max}$. All the potentials are embedded in an infinite square well potential, lying between $x=0$ and $x=a$. This plot is schematic only, and the $x$-axis is given in terms of the infinite square well width. In practice the potential would be provided with characteristic lengths, and {\em then} the infinite square well width would be chosen to ensure that states bound in the actual double well potential of interest would have negligible amplitude at the infinite wall boundaries (thus ensuring that the boundaries have no influence).}
\label{fig1}
\end{center}
\end{figure}

For curve-defined double wells we focus on two analytical forms; the first one consists of two parabolic wells,
\begin{equation}
V(x) = {V_{\rm cusp} \over b^2}\biggl( |x - \frac{a}{2}| - b\biggr)^2,
\label{potential}
\end{equation}
where $V_{\rm cusp} = {1 \over 2} m \omega^2 b^2$ is the value of the central maximum, defined here in terms of an oscillator frequency,
$\omega \equiv \sqrt{k/m}$,  where $m$ is the mass of the particle, and $k$ is the spring constant associated with the potential that exists on
either side of the central maximum. The central maximum is located at $x=a/2$, and the minima are symmetrically located at $x_{{\rm m}\pm} = a/2 \pm b$, as is
apparent in Fig.~\ref{fig1}. The other potential, which arises in many symmetry breaking transitions, is a quartic function of position,
\begin{equation}
V(x) = \frac{V_{\rm max}}{b^4} \biggl( \bigl( x - \frac{a}{2}\bigr)^2 - b^2 \biggr)^2,
\label{potential_2}
\end{equation}
where $V_{\rm max}$ is the value of the potential at the central maximum and $a/2 \pm b$ are the locations of the two local minima. In all cases the potential is symmetric about the point $x= a/2$, for reasons that will become clear below, and the two minima are centred around points $a/2 \pm b$.

In Ref. [\onlinecite{marsiglio09}] we demonstrated that a number of potentials could be solved through matrix mechanics, by expressing the wave function in terms of a set of basis functions, and formulating an eigenvalue problem. We did this using a very simple and (perhaps) the most familiar basis set known to undergraduate students at this stage, that of an infinite square well. Ref. [\onlinecite{marsiglio09}] solved the Schr\"odinger equation for familiar (but non-trivial) problems, such as the harmonic oscillator potential. This was done so the reader could readily understand the effects of the confining walls of an infinite square well potential, along with the necessary truncation required for a matrix which is in principle infinite; however, these truncation effects can be entirely (and systematically) eliminated by experimentation. That is, for the low-lying states, the infinite square well width can be made sufficiently large so the solutions are no longer sensitive to the walls; furthermore, the dimension of the matrix can be increased until convergence to any desired accuracy is attained.

In our formulation of the problem we require an infinite square well; in Fig.~1 this has been chosen to lie between $x=0$ and $x = a$ (also indicated with thick (red) vertical lines). The reason for placing this potential between $x=0$ and $x=a$ (rather than at $-a/2 < x < a/2$) is that the eigenstates of the infinite square well (without the double well potential) are more easily defined; they are conveniently given by
\begin{equation}
\psi_n(x) = \begin{cases}
\sqrt{\dfrac{2}{a}} \sin{\Big( \dfrac{n \pi x}{a} \Big)} & \text{if $0 < x < a$,} \\ 0 & \text{otherwise},
\end{cases}
\label{infinite_square_well_wavefunction} 
\end{equation}
with eigenvalues,
\begin{equation}
E_n^{(0)} = {n^2 \pi^2 \hbar^2 \over 2 m a^2} \equiv n^2E_1^{(0)}. 
\label{infinite_square_well_energies}
\end{equation}
The quantum number $n = 1,2,3, \ldots$ takes on a positive integer value, and we remind the reader that these states alternate between
even and odd around the midpoint, $x=a/2$.

Following Ref. [\onlinecite{marsiglio09}], we expand the solution to our full problem in
terms of this complete set:
\begin{equation}
| \psi \rangle = \sum_{m=1}^\infty c_m |\psi_m \rangle, 
\label{a1}
\end{equation}
and by inserting this into the Schr\"odinger equation and taking inner products with each basis state,
we eventually arrive at the eigenvalue equation
\begin{equation}
\sum_{m=1}^\infty H_{nm} c_m = E c_n, 
\label{a2} 
\end{equation}
where
\begin{eqnarray}
H_{nm} = & &\langle \psi_n|H|\psi_m \rangle = \delta_{nm}E_n^{(0)}\nonumber \\
& &+ {2 \over a}\!\int_0^a dx \sin{\Big(
\frac{n \pi x}{a} \Big)} V(x) \sin{\Big( {m \pi x \over a} \Big)} 
\label{ham_matrix}
\end{eqnarray}
is the Hamiltonian matrix, and $\delta_{nm}$ is the usual Kronecker delta function.

The integration in Eq.~(\ref{ham_matrix}) is elementary, regardless of which potential is used. Analytic expressions are provided in the Appendix, but one should keep in mind that these integrals can also be readily obtained numerically. In fact, herein lies one of the advantages for students to learn this method of solving the Schr\"odinger equation, as they can in principle solve for many realistic potentials.

Note that, just like in the harmonic oscillator problem, $H_{nm}$ is symmetric and peaked along the matrix diagonal. One might worry about the effects of truncation for this matrix problem; however, for large $n$, the diagonal elements increase as $n^2$. The off-diagonal elements quickly become negligible in comparison, and for large $n$ the energy levels approach those of an infinite square well, with an offset which accounts for the presence of the potential (the offset is equal to the average of the potential being considered).

\begin{figure*}
\begin{center}
\hspace{-2em}
\centerline{\includegraphics[height=3.5in,width=7.4in]{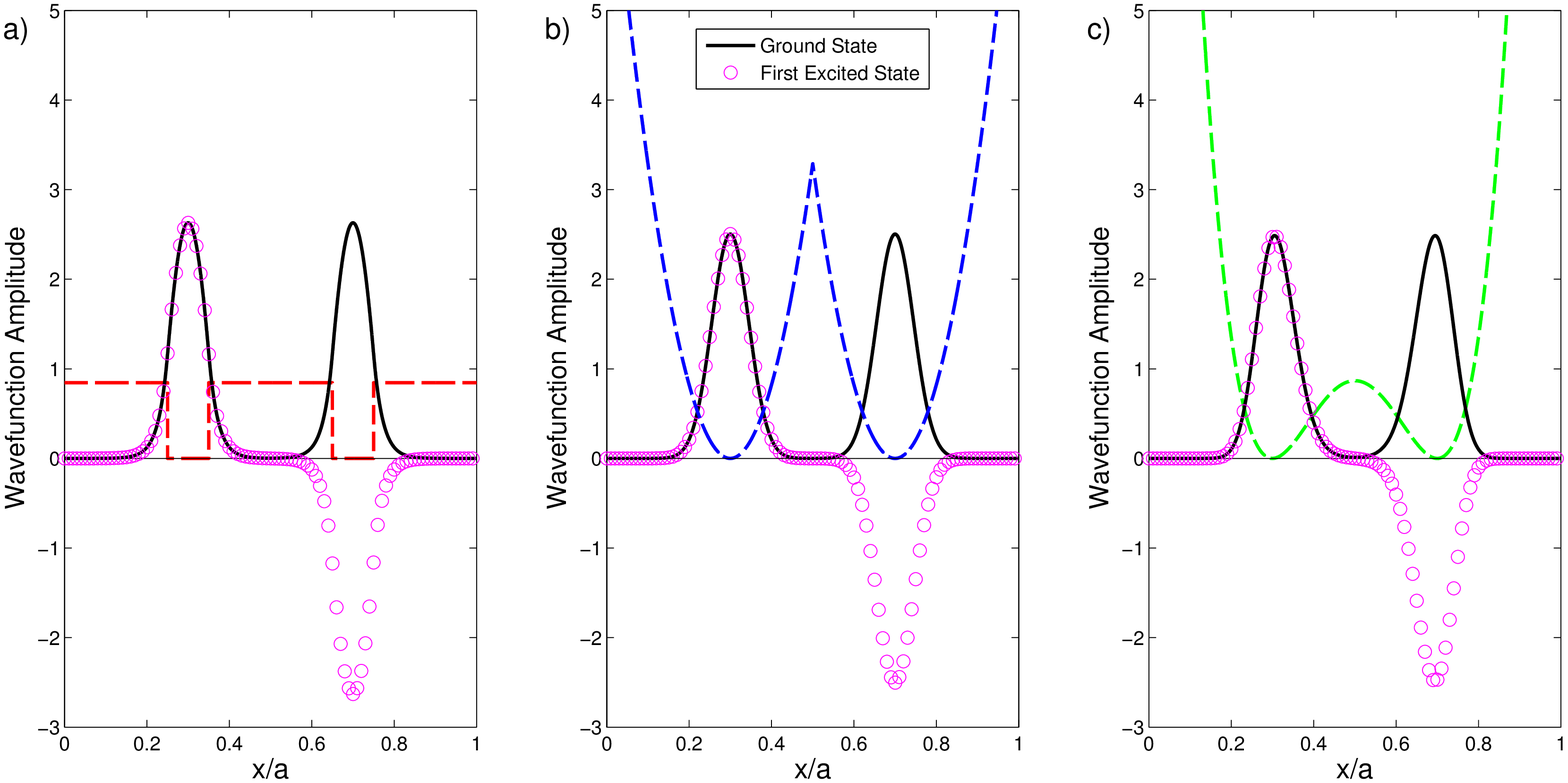}}
\caption{(color online) The ground state (solid black curves) and first excited state (purple circles) wavefunctions for the three potentials (shown as dashed red, blue, and green curves, for the square, quadratic, and quartic double well potentials, respectively) are shown. All potentials are symmetric about $x=a/2$ with minima located at $x_m=a/2 \pm b$, just as in Fig.~\ref{fig1}. A value of $b/a=0.2$ is used so that the infinite walls embedding the double well have no effect on the lower eigenstates; meanwhile, the height of the tunnelling barrier is such that several excited states occur in doubles (see Fig.~3 below). The values of $V_0$, $V_{\rm cusp}$, and $V_{\rm max}$ were determined by requiring the same ground state energy for all three wells. The square barrier potential (red curve in Fig. 2a) has an extra degree of freedom, the individual well widths, $c$. We used $c/a=0.1$ to make these results similar to those for the other two potentials. With a ground state energy, $E_1/E_1^{(0)} = 50$, $b/a=0.2$, (and $c/a=0.1$ for the square well), we determined $V_0/E_1^{(0)} \approx 253.6$, $V_{\rm cusp}/E_1^{(0)} \approx 987.0$, and $V_{\rm max}/E_1^{(0)} \approx 260.2$. Note that the wave functions for the ground state and the first excited state are essentially identical, except for the differing symmetry. Also, the three sets of wave functions are very similar to one another. The slightly narrower square double well result (a) is explained in the text.}
\label{fig2}
\end{center}
\end{figure*}

Equation~\eqref{a2}, with the use of equation~\eqref{mat_square}, \eqref{mat}, or \eqref{mat4} in the Appendix can be evaluated up to some cutoff for given parameters of the potential (for example, the parameters ${\hbar \omega/E_1^{(0)}}$ and $b/a$ in the case of the quadratic potential -- Eq.~\eqref{potential}). Equations~\eqref{mat_square}, \eqref{mat}, and \eqref{mat4} (each pertaining to a particular potential) form the elements of the matrix which is fed into an eigenvalue/eigenvector solver (see for example, Numerical Recipes\cite{press} or defined functions in software packages such as Matlab, Mathematica, or Maple). Typical results are shown in Fig.~\ref{fig2} for each of the potentials. To arrive at such results, students will have to truncate the matrix at some cutoff, $N$; a cutoff as low as $N=20$ leads to inaccuracies of less than $0.2\%$ for the ground state energies of the potentials \eqref{potential}, and \eqref{potential_2}, but the square well potential \eqref{square_doublewell_potential} requires a larger cutoff of  $N=50$ to be just as accurate. The reason for this is that the double square well potential presents a much narrower structure compared to the other two; therefore an accurate solution requires contributions from basis states with higher frequency components. These higher frequency states are only available at larger quantum number, $n$. By increasing the matrix size we can systematically produce the exact results to arbitrary precision.

There is a lot of room for experimentation; if one elects to study the two square wells, for example, students will discover that the convergence rate decreases dramatically as the well widths become narrower. This occurs for the same reason as it did in the case of a single square well,\cite{marsiglio09} namely that higher energy basis states are necessary not because they have higher energy, but, as explained above, because they oscillate in real space with shorter and shorter periods, and these more refined spatial variations are required to describe wave functions that are confined to narrower wells.

As one increases the maximum matrix size, $N$, the matrix solution will provide more eigenvalues and eigenvectors. 
It is important for students to realize that large eigenvalues will {\em not} correspond exclusively to a double well potential, but will also include the effects of the infinite square well in which the former is embedded. This means the higher eigenvalues and eigenvectors will closely resemble those pertaining to the infinite square well, as students can readily verify. Corrections to the infinite square well results correspond to accounting for the average of the double well potential, a result that students can readily verify for each potential.

Returning to Fig.~\ref{fig2}, we have chosen the potential parameters to obtain essentially the same ground state eigenvalue for each potential (see caption for details). In all cases the tunneling barrier that results is sufficiently large that the first excited state will be almost degenerate (slightly higher) with the ground state energy. We show the ground state and first excited state wave functions in each case. It is clear that they are generically the same; the ground state has even parity, and shows a large amplitude to be in either the left or right well. The first excited state has the same characteristics, except that it has odd parity. These results illustrate that the characteristics for low-lying states are qualitatively similar, independent of the precise form of the double well potential. Upon closer inspection it is also clear that the individual peaks of the wave function for the square double well potential (Fig.~2a) are slightly narrower than the other two. This is consistent with the fact that this well is more confining than the other two, and also with the fact noted earlier, that more basis states were required to obtain this solution accurately.

As mentioned above, one can continue with any (or all three) potentials and examine more characteristics of the tunneling between the two sides, or of the excited states. In the next section we focus on one particular potential, the one which is quadratic in $x$, given by Eq.~\eqref{potential}, and examine the eigenvalues more closely. Results for this potential are shown in Fig.~\ref{fig3}. Focusing on just the numerical results, note that as one moves up in energy, we pass through the different regimes; at low energies the levels are split and come in almost degenerate pairs, while at higher energies the levels spread out, corresponding to a regime where the central region is no longer forbidden. Finally, at very high energies the spectrum begins to approach quadratic (in $n$) behaviour, corresponding to an infinite square well potential. In the next section we further discuss these results after providing a brief summary of the WKB results.

\section{The double well potential: WKB approximation}

As indicated above, there are three distinct regions in this problem; these regions are not completely obvious from the numerical solution, but their presence needs to be explicitly accounted for when undertaking a WKB calculation. The reader should note that the WKB solution for a double well potential is presented in most texts,\cite{landau} but usually the treatment is confined to a calculation of the splitting between the two lowest energy levels. This (tiny) splitting is connected to the tunneling probability between the left and right side wells, giving rise to issues analogous to those that arise in the case of the Schr\"odinger cat.\cite{maris00} The WKB approximation for all three regions differs in details -- depending on whether the energy level is below the central maximum (region I), $E < V(a/2)\equiv
V_{\rm cusp}$, above the central maximum but below the value of the double well potential at the walls (region II), $V_{\rm cusp} < E < V(a)$, or above this value (region III), $E > V(a)$.

It is worthwhile, given that numerically exact solutions are so readily available with this method, to investigate how accurate the WKB solution really is. The solutions physically relevant to a double well all lie in region I; beyond this our method of solution (i.e. the existence of infinite walls) affects the numerical solutions.\cite{remark3} Nonetheless once one works out the WKB solution in region I, the superb agreement with the numerically exact results encouraged us to investigate the degree of agreement in the other two regions. As we see below, the agreement is excellent there as well, and we therefore describe each of these in the subsections below. These calculations are slightly more involved; the undergraduate student may want to follow the solution in region I, while a full survey of all three regions may be more appropriate for students working on specialized projects (related to the WKB approximation), or for graduate students. 

\subsection{region I}

\begin{figure*}
\begin{center}
\hspace{-3.9em}
\includegraphics[height=6.5in,width=5.0in, angle = -90]{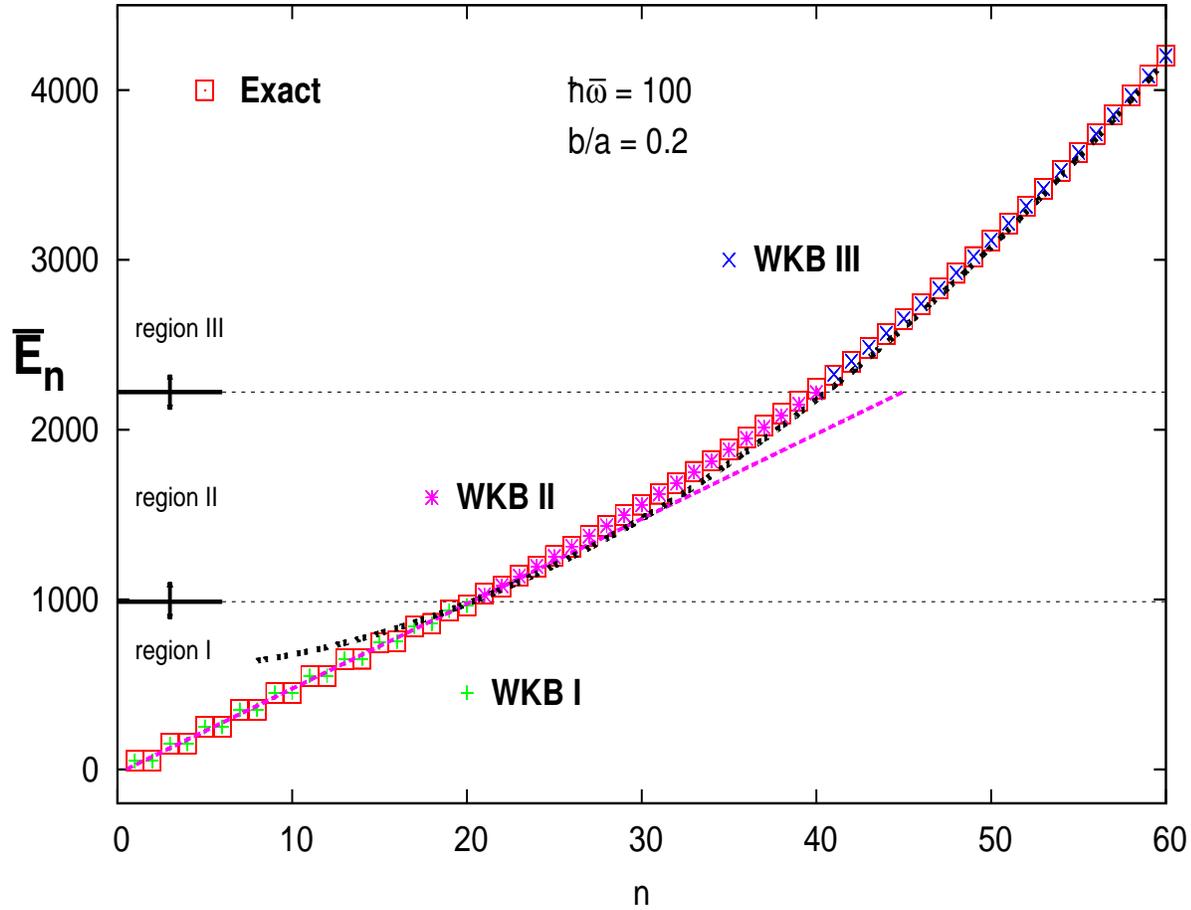} 
\caption{(color online) Eigenvalues obtained by the numerical diagonalization of the double well potential~\eqref{potential} embedded in an infinite square well (see Fig.~\ref{fig1}). The numerically exact results are shown with red squares (the actual points are at the center of the squares). We used a truncation size of $N=200$ to obtain these results, but a much smaller size would suffice as well, especially for the low lying eigenvalues. We also show with various symbols the WKB approximations appropriate to the three different regions, as discussed in the text (they are also color coded here).
They are remarkably accurate on this energy scale. Also shown is a dashed (pink) straight line starting from near the origin; it is given by $\bar{E}_n = (\hbar \bar{\omega}/2)(n - 1/2)$, which represents the result expected for a single harmonic oscillator potential. This expectation is relevant when the two parts of the double well are `very far away' from one another, and so the two parts of the double well appear to behave independently. It is clear that the levels in Region I satisfy this criterion very well. (Note: the factor of 2 (in $\hbar \bar{\omega}/2$) is required because there are two wells, and for the same reason we need twice as many quantum numbers. One should think of the pairs of values in region I as sharing a single quantum number which is the average of the two.) Also shown as the black double-dotted curve is the analytic expression, given by Eq.~\eqref{offset}; this was derived through first order perturbation theory and is expected to be very accurate for the higher levels in Region III. Indeed, it is already fairly accurate at $n=60$.}
\label{fig3}
\end{center}
\end{figure*}

Region I (below the central maximum) is the region treated in most textbooks and pedagogical papers, since the splitting in energy levels that results from tunnelling is the most important outcome of the double well potential. While textbooks often deal with the ground state,\cite{landau} we wish to describe not just the ground state, but all eigenstates whose energy lies in Region I. Thus, following the methodology of Ref. [\onlinecite{landau}], but avoiding their approximations, leads to the following transcendental equation,
\begin{equation}
\bar{E}_{n^\prime \pm} = \hbar \bar{\omega} \bigl(n^\prime-{1 \over 2}\bigr) \pm {\hbar \bar{\omega} \over \pi} {\rm tan}^{-1}\biggl({e^{-\phi_{n^\prime \pm}} \over 2}\biggr),
\label{wkb1}
\end{equation}
where $\bar{Q} \equiv Q/E_1^{(0)}$ for all energies, and the quantity $\phi_{n^\prime}$ is itself a function of the energy that we are seeking:
\begin{eqnarray}
&& \phi_{n^\prime} = {2 \bar{V}(\frac{a}{2}) \over \hbar \bar{\omega}} \Biggl(\sqrt{1 - {\bar{E}_{n^\prime \pm} \over \bar{V}(\frac{a}{2})}} - {\bar{E}_{n^\prime \pm} \over \bar{V}(\frac{a}{2})}\Biggl[ {1 \over 2} 
\ln{ \left( {\bar{V}(\frac{a}{2}) \over \bar{E}_{n^\prime \pm}}\right) } \nonumber \\
&&+ \ln{\left(1 + \sqrt{1 - {\bar{E}_{n^\prime \pm} \over \bar{V}(\frac{a}{2})}} \right)} \Biggr] \Biggr).
\label{phi}
\end{eqnarray}
Eqs. (\ref{wkb1},\ref{phi}) must be solved iteratively. Note that there is an integer index $n^\prime$ which is associated with two (split) levels, and we reserve the  label $n$ (which appears in Fig.~3, for example) to enumerate all the eigenvalues. In practice the iteration converges very quickly, as in the ground state, for example, $\bar{E} << \bar{V}(a/2)$, so $\phi \approx {2 \bar{V}(a/2) \over \hbar \bar{\omega}}$, as derived in Ref. [\onlinecite{landau}], and Eq. (\ref{wkb1}) provides a closed form solution for the two energies. Similarly, the argument of the inverse tangent function is very small, so we obtain for the ground state, in the case of a relatively high central barrier,
\begin{equation}
\Delta \equiv E_+ - E_- \approx {\hbar \omega \over \pi} e^{-\phi},
\label{delta}
\end{equation}
where
\begin{equation}
\phi \approx {2 {V}(a/2) \over \hbar {\omega}} = {\pi^2 \over 2}\biggl({\hbar \omega \over E_1^{(0)}} \biggr) \biggl({b \over a}\biggr)^2,
\label{delta2}
\end{equation}
all of which is well known.\cite{landau}
 
The numerical results (iterated solutions to Eqs. (\ref{wkb1},\ref{phi})) are shown in Fig.~3 as green cross-hairs (inside the squares in Region I), and clearly coincide with the (numerically) exact results (the centres of the squares). Of course the splitting is a much finer detail and will be examined further below. Note that the 1D quantum harmonic oscillator result is also shown (dashed pink line starting near the origin); the numerically exact and WKB results nicely follow that of a single harmonic oscillator for most of Region I.

\subsection{region II}

Region II involves intermediate energies, as indicated in Fig.~3. The WKB procedure is straightforward, and results in the transcendental equation:
\begin{equation}
\bar{E}_n = { \hbar \bar{\omega} \bigl( n - {1 \over 2} \bigr) \over 1 + {2 \over \pi} \biggl[ \sqrt{\bar{V}({a \over 2}) \over \bar{E}_n} \sqrt{ 1 - {\bar{V}({a \over 2}) \over \bar{E}_n}} + {\rm sin}^{-1} \sqrt{\bar{V}({a \over 2}) \over \bar{E}_n} \biggr] }
\label{wkb2}
\end{equation}
where the `bar' over each energy indicates it is normalized to the infinite square well ground state energy, $E_1^{(0)}$. Once again, this equation requires iteration,
but we find convergence is readily attained within a handful of iterations. The results are also plotted in Fig.~3 as pink asterisks (in the squares in Region II). In this region there is a distinct departure from the result describing a single harmonic oscillator, but the WKB results are in very good agreement with the numerically exact ones.

\subsection{region III}

Region III consists of energy levels above the point where the double well potential is cut off by the walls. This region exists only because of our method of numerically exact solution, but we include a comparison nonetheless. The WKB approximation is now specific to that involving two vertical walls. The equation governing
this result is
\begin{equation}
\bar{E}_n = {n^2 \over \bigl[1 + \delta_1(a) + \delta_1({a \over 2}) + \delta_2(a) + \delta_2({a\over 2})\bigr]^2}
\label{wkb3}
\end{equation}
where the corrections labelled $\delta_i$, $i = 1,2$, tend to be small, but are dependent on the energy being sought. These corrections are given by
\begin{eqnarray}
\delta_1(x) &=& {2 \over \pi} {1 \over \hbar \bar{\omega}} \sqrt{\bar{V}(x)} \biggl( \sqrt{1 - {\bar{V}(x) \over \bar{E}_n}} - 1 \biggr), \nonumber \\
\delta_2(x) &=& {2 \over \pi} {\sqrt{\bar{E}_n} \over \hbar \bar{\omega}} \biggl( {\rm sin}^{-1} \sqrt{\bar{V}(x) \over \bar{E}_n} - \sqrt{\bar{V}(x) \over \bar{E}_n} \biggr).
\label{corr}
\end{eqnarray}
Eqs. (\ref{wkb3}) and (\ref{corr}) are iterated to convergence with very little difficulty. These results are shown as the (blue) $x$'s (in the squares in Region III), and are in excellent agreement with the numerically exact results. In fact, a good starting approximation is obtained by using first order perturbation theory (with the double well potential as the perturbation and particle in the infinite square well as the unperturbed Hamiltonian), and one obtains\cite{marsiglio09}
\begin{equation}
{E_n \over E_1^{(0)}} \approx n^2 + {\pi^2 \over 48} \biggl( {\hbar \omega \over E_1^{(0)}}\biggr)^2 \biggl( 1 -6 {b \over a} + 12 \Bigl({b \over a}\Bigr)^2  \biggr).
\label{offset}
\end{equation}
This curve is also shown in Fig.~\ref{fig3} as a double-dotted (black) curve, and, though not as accurate as the WKB approximation, works fairly well for levels in Region III.

These results indicate the remarkable accuracy of the WKB approximation, not just for the ground state, but for the excited states as well. The latter is not too surprising, since the WKB is a semiclassical approximation, and ought to improve as the quantum number increases; however, this is not often emphasized in comparisons of this sort.

\subsection{The energy splitting}

While the comparison of the WKB approximation to the numerically exact results in Fig.~3 may look impressive, the scale there is fairly coarse; indeed a `large' error in the splitting of the levels in Region I would go unnoticed on such a plot. Therefore we take a closer look at this energy splitting, since, ultimately, it is this splitting which plays a primary role in the tunnelling phenomenon.

\begin{figure}[h!]
\begin{center}
\includegraphics[height=3.5in,width=3.0in, angle = -90]{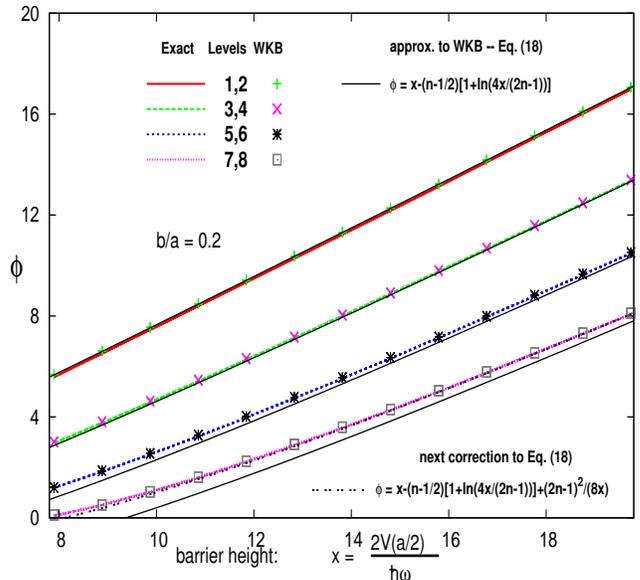} 
\caption{(color online) Plot of the exponent $\phi$ using the WKB approximation (Eqs.~\eqref{wkb1} and \eqref{phi} -- symbols) and as defined by the numerically exact results (Eq.~\eqref{phi_exact} -- colored curves). The WKB approximation is in excellent agreement with the numerically exact solutions for the four pairs of split states, even down to the point where the barrier is about to disappear for the 4th pair of  states (lower left). Also shown are analytical approximations for the WKB approximation as discussed in the text; these are plotted as thin solid (black) curves, and are given by Eq. (\ref{phi_approx1}). The double-dotted (black) curve includes a further correction discussed below Eq. (\ref{phi_approx1}), and is included only for the 4th pair of states, as it is not needed for the others. These analytic approximations to the WKB iterated solution work very well for all the pairs of states shown.}
\label{fig4}
\end{center}
\end{figure}

The WKB approximation motivates a suitable definition for the energy splitting observed for the low lying states. Using Eq. (\ref{wkb1}) as a guide, we define the energy splitting to be $\Delta_{n^\prime} \equiv E_{n^\prime+} - E_{n^\prime-}$, where the index $n^\prime$ refers to the pair of eigenvalues clearly discernible in Fig.~\ref{fig3}. A comparison of these energy splittings with the WKB approximation spans several decades, so we find it more convenient to instead focus on the exponent $\phi$ that appears in Eq. (\ref{wkb1}).  Within the WKB approximation this is defined by the iterated result of Eq. (\ref{phi}) (in combination with Eq. (\ref{wkb1})), and we will henceforth refer to this as $\phi_{n^\prime {\rm WKB}}$. On the other hand, the exact result is {\em defined} by

\begin{equation}
\phi_{n^\prime \rm ex} \equiv - \ln{\biggl[2 \  {\rm tan}\biggl( {\pi \over 2}{\bar{E}_{n^\prime +} - \bar{E}_{n^\prime -} \over \hbar \bar{\omega}} \biggr)\biggr]},
\label{phi_exact}
\end{equation}

\noindent
where the $\bar{E}_{n^\prime \pm}$ are obtained numerically. The numerically exact results are compared with the WKB in Fig.~\ref{fig4}; the exponent $\phi$ is shown plotted as a function of $2V(a/2)/(\hbar \omega) \equiv (\pi^2/2)\hbar \bar{\omega} (b/a)^2$. This is the barrier height with respect to the approximate ground state energy level. In Fig.~\ref{fig4} this ratio is changed by varying $\omega$ such that $40 < \hbar \bar{\omega} < 100$. Clearly the agreement is (very) excellent over the entire range of $\hbar \bar{\omega}$ shown, for the first 4 pairs of energy levels. Note that this is true even at the lower range of {\it x} (i.e. $\omega$ just below the central maximum), where the 4th pair is just about to enter region II, so this is the limit of essentially no barrier at all. Closer scrutiny shows that the WKB approximation very slightly overestimates the level of $\phi$ and therefore slightly underestimates the energy splitting, compared to the numerically exact result. Moreover, we obtain a very good estimate for $\phi_{n^\prime \rm{WKB}}$ by expanding Eq. (\ref{phi}) assuming that $\bar{E}_{n^\prime \pm} \approx \hbar \bar{\omega}(n^\prime - 1/2)$, and that $x \equiv 2V(a/2)/(\hbar \omega)$ (the barrier height in Fig.~4) is large; the result is

\begin{equation}
\phi \approx x - \biggl(n^\prime-{1 \over 2}\biggr)\biggl[ 1 + \ln{\biggl({4x \over 2n - 1}\biggr)}\biggr],
\label{phi_approx1}
\end{equation}
and these are also shown in Fig.~4 (solid black curves). This expression works very well for the ground state, but begins to deteriorate for the excited states, especially at the lower range of $x$ ($\omega$ near the central maximum). Expanding to the next order adds a term ${(2n-1)^2 \over 8x}$ to Eq. (\ref{phi_approx1}), which is shown for the 4th pair only as a black double-dotted curve --- this correction essentially restores the approximate value to the fully iterated WKB value and even more so for the other three pairs (not shown).
We conclude that the WKB approximation is indeed very accurate, even when examining finer energy scales associated with the splitting of the low-lying levels that results
from a very small tunneling amplitude between the two wells.

\section{Summary}

In this paper we have provided an illustration of the numerical solution to a non-trivial problem, the double well potential. The focus has {\em not} been on the numerical implementation; on the contrary, we have used this example in the classroom, with the expectation that students have software tools (Mathematica, Maple, MatLab, etc.) readily available. In practice this is not always the case, but our belief is that this expectation is slowly becoming an integral part of an undergraduate education in science or engineering. Assuming this is so, then this method of solution requires a conceptual understanding of mathematical constructs that are typically taught in the first year of university: integrals in calculus and linear algebra. We emphasize that {\em conceptual understanding} only is required, since implementation of both of these concepts is readily accomplished through software tools. The alternative is to first understand the (conceptual) solution to differential equations, and then utilize numerical software to implement these solutions; this is a much more difficult task, since an understanding of differential equations lags behind introductory calculus and linear algebra by about two years. In our experience with teaching\cite{remark4} Quantum Mechanics, much of the students' understanding of differential equations emerges from advanced physics courses (such as QM). This in turn focuses attention away from the physical systems and onto understanding the mathematics, which is precisely what we hope to avoid with the present approach.

We have emphasized the characteristics that arise in the solution of the double well potential that one can teach conceptually through perturbation or variational procedures. However, by using the (numerically exact) method presented here, the student gains several advantages. First, the solution is numerically exact, so lingering doubt about the validity of the approximate solutions is no longer possible. Second, this method of solution can be applied to any problem of this kind, so the potential doesn't even have to be analytically defined. Third, this method of solution is an implementation of what is usually taught formally in most undergraduate quantum mechanics courses, and certainly reinforces the more abstract concepts of Hilbert space, etc. Finally, this is precisely the method of solution used in research (at least in condensed matter problems, one rarely solves a differential equation --- usually problems are formulated in terms of matrices). Therefore practice with these methods can better equip a student for research in physics.

It is clear from the double well examples presented, that for the low-lying states, the solutions have much in common. The wave function amplitude is peaked around each of the wells, so classically speaking, the particle can be viewed as being in a superposition of two states; in the first the particle is in the left well, while in the second the particle is in the right well. With respect to these solutions, it is clear by visual inspection of Fig.~\ref{fig2}, for example, that the vertical walls that constitute the embedding infinite square well potential have {\em no influence} on the solutions, as all wave functions shown have decayed to zero at the walls.

In addition we have carried out a detailed analysis of the WKB approximation for a quadratic double well potential, as this method is often used for such problems in real-world applications. We have determined the WKB energy levels self-consistently, and found that this approximation is remarkably accurate, not just for the ground state, but for the excited states as well.

\begin{acknowledgments}

This work was supported in part by the Natural Sciences and Engineering Research Council of Canada (NSERC), and by the Teaching and Learning Enhancement Fund (TLEF) at the University of Alberta.

\end{acknowledgments}

\appendix

\section{matrix elements}

As remarked in the text, since it is assumed that the student is using a software package to diagonalize the 
matrix, it is likely straightforward to utilize the same package to evaluate the matrix elements either analytically or
numerically. Alternatively, to proceed `by hand,' we use the identity
\begin{equation}
2\sin{a} \sin{b} = \cos{(a-b)} - \cos{(a+b)},
\label{sin_identity}
\end{equation}
to simplify the integral in equation~\eqref{ham_matrix}.

For the square double well, the potential is non-zero and constant ($V_0$) over the intervals
\begin{eqnarray}
0< & x & < {a \over 2} - b - {c \over 2} \nonumber \\
{a \over 2} - b + {c \over 2} < & x & < {a \over 2} + b - {c \over 2} \nonumber \\
{a \over 2} + b + {c \over 2} < & x & < a,
\label{intervals}
\end{eqnarray}
so it is a simply a matter of integrating $\cos{(x)}$ over these intervals. The result is (with $E_1^{(0)}$ as the unit of energy)
\begin{flalign}
&{(h_s)}_{nm} \equiv  {(H_s)}_{nm}/E_1^{(0)} = \delta_{nm} \biggr[ n^2 & \nonumber
\end{flalign}
\\[-24pt] 
\begin{eqnarray}
& + & {V_0 \over E_1^{(0)}}\biggl( 1 - {2c \over a} + {2(-1)^n \over \pi n}\cos{\left(\frac{2 \pi n b}{a}\right)} \sin{\left(\frac{\pi n c}{a}\right)} \biggr) \biggr] \nonumber \\
& + &\bigl({1 - \delta_{nm}}\bigr) {4 \over \pi}{V_0 \over E_1^{(0)}}\biggl( q(n+m) - q(n-m) \biggr),
\label{mat_square}
\end{eqnarray}
where
\begin{equation}
q(n) \equiv {1 \over n} \cos{\biggl({\pi n \over 2}\biggr)} \cos{\biggl({\pi n b \over a}\biggr)} \sin{\biggl({\pi n c \over 2a}\biggr)}.
\label{qofn}
\end{equation}
We have used the subscript `s' to denote the `square' double well potential.

For the quadratic and quartic potentials
only integrals with even powers of $x$ times $\cos{x}$ are required:
\begin{equation}
I_j(n) \equiv \int_0^{1/2} \ dx \ x^j  \cos{(n \pi x)},
\label{cos_int}
\end{equation}
where the integration need only extend to the half-way point because of the symmetry of the potential.
The result for the quadratic potential \eqref{potential} is
\begin{eqnarray}
&&{(h_2)}_{nm} \equiv  {{(H_2)}_{nm} \over E_1^{(0)}} = \delta_{nm} \bigg[n^2 +{\pi^2 \over 48} \biggl({\hbar \omega \over E_1^{(0)}}\biggr)^2 f_n\biggl({1 \over 2} - {b \over a}\biggr) \bigg] \nonumber \\
& & + \bigl({1 - \delta_{nm}}\bigr) \biggl[ {1 + (-1)^{n+m} \over 2} \biggl({\hbar \omega \over E_1^{(0)}}\biggr)^2 g_{nm}\biggl({1 \over 2} - {b \over a}\biggr) \biggr],
\label{mat}
\end{eqnarray}
where
\begin{flalign}
&f_n(x) = 1 - {6(-1)^n \over (\pi n)^2} -6x\biggl[1 - 2{(-1)^n - 1 \over (\pi n)^2} \biggr] + 12x^2,&
\label{f}
\end{flalign}
and
\begin{flalign}
&g_{nm}(x) = {1 \over 2} \biggl( {(-1)^{(n-m)/2} \over (n-m)^2} - {(-1)^{(n+m)/2} \over (n+m)^2} \biggr)& \nonumber
\end{flalign}
\\[-26pt] 
\begin{equation}
- x \biggl( {(-1)^{(n-m)/2} - 1 \over (n-m)^2} - {(-1)^{(n+m)/2} - 1 \over (n+m)^2} \biggr).
\label{g}
\end{equation} 

We can formulate an equivalent expression to equation~\eqref{mat}, where if $n$ and $m$ are both odd,
\begin{flalign}
&{(h_2)}_{nm} \equiv  {(H_2)}_{nm}/E_1^{(0)} = & \nonumber
\end{flalign}
\\[-30pt] 
\begin{equation}
\frac{V_{cusp}}{E_1^{(0)}}\frac{8 m n \bigl[1 - \delta \bigl(2+\bigl({m \over n}+{n \over m}\bigr)(-1)^{{n+m \over 2}} \bigr) \bigr]}{\bigl[ \delta \pi (m-n) (m+n) \bigr]^2},
\label{mat_square_2a}
\end{equation}
where $\delta \equiv b/a$, and if $n$ and $m$ are both even,
\begin{flalign}
&{(h_2)}_{nm} \equiv  {(H_2)}_{nm}/E_1^{(0)} = & \nonumber
\end{flalign}
\\[-30pt] 
\begin{equation}
\frac{V_{cusp}}{E_1^{(0)}}\frac{8 m n \bigl[1- 2 \delta \bigl(1-(-1)^{{n+m \over 2}} \bigr) \bigr]}{\bigl[ \delta \pi (m-n) (m+n) \bigr]^2},
\label{mat_square_2b}
\end{equation}
and when $n = m$,
\begin{flalign}
&{(h_2)}_{nn} \equiv  {(H_2)}_{nn}/E_1^{(0)} = n^2 + \frac{V_{cusp}}{E_1^{(0)}} \biggl[1 & \nonumber
\end{flalign}
\\[-30pt] 
\begin{equation}
+\frac{1}{\delta}\left(\frac{1-(-1)^n}{(n \pi)^2}-\frac{1}{2}\right) + \frac{1}{\delta^2} \left( \frac{1}{12}-\frac{1}{2 (n \pi)^2} \right) \biggr],
\label{mat_square_2c}
\end{equation}
and for all other $n$ and $m$, ${(h_2)}_{nm}=0$.

Here we have used the subscript `2' to denote the quadratic potential. For equation~\eqref{mat} we have adopted the (dimensionless) energy ratio, ${\hbar \omega \over E_1^{(0)}}$, to denote the strength of the potential (recall $V_{\rm cusp} \equiv {1 \over 2} m \omega^2 b^2$, where $V_{\rm cusp}$ is the value of the central maximum and the locations of the minima at $x_m = a/2 \pm b$ defines the parameter $b$). Also note that, just like in the harmonic oscillator problem, the $g_{nm}$~\eqref{g} remain of order unity close to the diagonal. Nonetheless, as in that problem, for large $n$, the diagonal elements grow as $n^2$, so the off-diagonal elements become negligible in comparison, and the energy levels approach those of an infinite square well, with an offset to account for the double well potential. This is done by averaging the potential and adding the average to the upper eigenvalues of the infinite square well, thus producing shifted infinite well energies for the upper eigenvalues of the embedded double well potential.

Finally, for the quartic potential~\eqref{potential_2},  using the identity~\eqref{sin_identity} and the substitution $y = x/a - 1/2$ results in the expression,
\begin{flalign}
&{(h_4)}_{nm} \equiv  {(H_4)}_{nm}/E_1^{(0)} = \delta_{nm} n^2 + \cos{\left({\pi (n+m)\over 2}\right)}& \nonumber
\end{flalign}
\\[-30pt] 
\begin{equation}
\times {2 \over \delta^4}{V_{\rm max} \over E_1^{(0)}} \bigl[ (-1)^m J(n-m) - J(n+m) \bigr],
\label{mat4}
\end{equation}
where $\delta \equiv b/a$, and
\begin{equation}
J(n) \equiv I_4(n) - 2\delta^2 I_2(n) + \delta^4 I_0(n),
\label{jk}
\end{equation}
where $I_j(n)$ is defined in equation~\eqref{cos_int}.
These integrals are readily obtained through integration by parts, integral tables, or analytic integral solvers such as Maple and Mathematica. The required expressions are
\begin{equation}
I_0(n) = {1 \over 2} \delta_{n,0} + (1 - \delta_{n,0}){1 \over \pi n} \sin{\left(\pi n\over 2\right)},
\label{i0}
\end{equation}
\begin{eqnarray}
&&I_2(n) = {1 \over 24} \delta_{n,0} + (1 - \delta_{n,0})\times \nonumber \\
&& \biggl[ \left({1 \over 4\pi n} -{2 \over (\pi n)^3} \right) \sin{\left(\pi n \over 2\right)} + \frac{\cos{\left(\pi n\over 2\right)}}{(\pi n)^2} \biggr],
\label{i2}
\end{eqnarray}
and
\begin{eqnarray}
I_4(n) &=& {1 \over 160} \delta_{n,0} + (1 - \delta_{n,0})\times \nonumber \\
&& \biggl[ \biggl({1 \over 16\pi n} -{3 \over (\pi n)^3} + {24 \over (\pi n)^5} \biggr) \sin{\left(\pi n \over 2\right)} \nonumber \\
&& +\biggl({1 \over 2(\pi n)^2} - {12 \over (\pi n)^4} \biggr) \cos{\left(\pi n\over 2\right)} \biggr].
\label{i4}
\end{eqnarray}

Alternatively, the following equivalent expression can be obtained using analytic integration solvers such Maple and Mathematica (with appropriate simplifying assumptions such as $n$ and $m$ being integers):
\begin{flalign}
&{(h_4)}_{nm} \equiv  {(H_4)}_{nm}/E_1^{(0)} = {\delta}_{nm} \biggl[n^2 +\frac{V_{max}}{E_1^{(0)}}\biggl(1 & \nonumber
\end{flalign}
\\[-30pt] 
\begin{equation}
-\frac{1}{\delta^2}\left( {1 \over 6}-\frac {1}{(n \pi)^2} \right) + \frac{1}{\delta^4} \left( \frac{1}{80} - \frac{1}{4(n \pi)^2} + \frac{3}{2 (n \pi)^4} \right) \biggr] \nonumber
\end{equation}
\\[-30pt] 
\begin{equation}
+(1-{\delta}_{nm})\frac{V_{max}}{E_1^{(0)}}\biggg[\frac{2mn \left( 1+ \left( -1 \right) ^{n+m} \right)}{\bigl[\pi \delta \left( n-m \right) \left( n+m \right)\bigr]^2} \nonumber
\end{equation}
\\[-30pt] 
\begin{equation}
\times \left( {\frac {{\pi }^{2} \left( {n}^{2}-{m}^{2} \right) ^{2}-48(n^2+m^2)}{\bigl[\pi \delta \left( n-m \right) \left( n+m \right)\bigr]^2}}-4 \right) \biggg].
\label{mat4_2}
\end{equation}

As stated in the text, these can be readily checked by a simple numerical evaluation of the original integral.


\begin{thebibliography}{1}
\bibitem{remark2} We emphasize that an undergraduate student that has completed his/her first year of study can technically carry out all the calculations. Only a knowledge of elementary calculus, linear algebra, and the ability to use a software package like Mathematica, Maple, or Matlab is required. This being said, of course the physical concepts discussed here are suitable for students in their final year of undergraduate studies. In contrast, the WKB calculations presented here are more
difficult.
\bibitem{remark4} In fact one of us (FM) has been teaching Quantum Mechanics, while the other (VJ) has been learning it, so this is a shared perspective.
\end{thebibliography}
\end{document}